\documentclass[pre,twocolumn,showpacs,aps,floats,floatfix]{revtex4}

\usepackage{epsfig}
\usepackage{epstopdf}
\usepackage{amssymb,amsmath}

\begin{document}

\title{Electrostatic self-energy of a partially formed spherical shell in salt solution: \\
application to stability of tethered and fluid shells -- viruses and vesicles}

\author{An\v{z}e Lo\v{s}dorfer Bo\v{z}i\v{c}\footnotemark[1], Antonio \v{S}iber\footnotemark[2], and Rudolf Podgornik\footnotemark[1]\footnotemark[3]}
\affiliation{\footnotemark[1]Department of Theoretical Physics, Jo\v{z}ef Stefan Institute, SI-1000 Ljubljana, Slovenia\\\footnotemark[2]Institute of Physics, Bijeni\v{c}ka cesta 46, P.O. Box 304, 10001 Zagreb, Croatia\\\footnotemark[3]Department of Physics, Faculty of Mathematics and Physics and Institute of Biophysics, School of Medicine, University of Ljubljana, SI-1000 Ljubljana, Slovenia}

\begin{abstract}
We investigate the electrostatics of a partially formed, charged spherical shell in a salt solution. We solve the problem numerically at the Poisson-Boltzmann level and analytically in the Debye-H\"{u}ckel regime. From the results on energetics of partially formed shells we examine the 
stability of tethered (crystalline) and fluid shells towards rupture. We delineate different regimes of stability, where, for fluid shells, we also include the effects of bending elasticity of the shells. Our analysis shows how charging of the shell induces its instability towards rupture but also provides insight regarding growth of charged shells.
\end{abstract}
\pacs{87.15.Fh,41.20.Cv,87.15.Pc}
\date{\today}
\maketitle

\section{Introduction}

Self-assembly as well as disassembly of biological macromolecules and their aggregates is governed by the
standard principles of colloid and nanoscale stability theory \cite{French}: repulsive interactions of in general disparate origins are usually counteracted by the ubiquitous van der Waals (vdW) force \cite{Parsegian}. For macromolecular aggregates bearing net charges, the Coulomb interaction in one of its many guises \cite{Wenbing,LesHouches} is the leading component of repulsive interactions, while for those decorated with exposed hydrophobic moieties the hydrophobic interaction \cite{Chandler}, at least in part related to the vdW interactions, joins in the overall attractive part of the stability condition. A stable state of (classical) matter is possible only if other forces apart from electrostatic ones are present \cite{Stratton}. In addition it was shown recently \cite{Rahi} that similarly the vdW interactions alone, at least for bodies with permittivities that are all 
higher or lower than that of the intervening medium, are also not able to generate stable states of matter. Only by combining the two together, or more unusually having bodies with alternating higher/lower permeabilities than the intervening medium, can full stability of matter be achieved.

In biological context it is mostly the screened electrostatic interactions and hydrophobic/van der Waals interactions that set the structural stability conditions of macromolecular aggregates \cite{LesHouches}. 
Of the two mentioned above, the hydrophobic force corresponds at least in part to an effective interaction arising  from the statistical properties of water molecules around the hydrophobic moieties \cite{Tanford}, while the van der Waals force is a  true force which can be modified by the presence of the aqueous solvent but acts also in its absence. If the attractive forces are strong enough to overcome the electrostatic repulsion, the macromolecules that in general contain charged as well as hydrophobic patches may cluster into different stable aggregate structures, depending on their geometry as well as on the spatial distribution of hydrophobic and hydrophilic patches along their surface. Since the presence of an aqueous solution, which in the biological milieu contains various dissolved ionic species, affects mostly the properties of the electrostatic interaction via a pronounced Debye-H\" uckel screening, the process of aggregation itself is also influenced by the screening properties of the solvent, i.e. the concentration and valency of salt ions. 

An interesting example of a soft matter system where the electrostatic interactions determine its stability is the problem of the shape of a charged fluid droplet studied by Lord Rayleigh more than a century ago \cite{Rayleigh}. He showed that a charged droplet has a limited range of stability and breaks into smaller droplets if its charge exceeds a certain critical value. This critical value of the charge is set by the balance between its electrostatic energy, proportional to the square of the charge, and its surface energy, proportional to the surface tension of the air-water interface. The latter can be associated with hydrophobic interactions at the fluid-gas interface. Beyond the limit of stability set by the critical value of the charge, the equilibrium configuration is a collection of smaller droplets with charges below the critical value, placed at an infinite distance from each other.

The concept of Rayleigh's instability in charged droplets described above can also be applied to polyelectrolyte globules in poor solvents, which again implies the existence of hydrophobic attractions between monomers. The stability limit in this case depends on the fraction of charged monomers in the polyelectrolyte chain.  Just as in the case of a charged droplet, a polyelectrolyte chain in a poor solvent can reduce its energy by splitting into a set of smaller charged globules. Because of the connectivity of the polymer chain, the smaller globules have to remain connected by an intervening finite polymer segments \cite{Rubinstein}. In such a way the equilibrium structure after the limit of stability of the original polymer globule is similar to a pearl necklace. Beyond the limit of stability the polyelectrolyte chain first takes a dumbbell configuration, turning into a pearl necklace with three beads joined by two segments, leading eventually to a whole cascade of transitions between necklaces with different number of beads as the charge on the chain increases \cite{Rubinsteinreview}.

The system that we study in this paper belongs to the class of systems discussed above. In what follows, we will be interested in stability of spherical, shell-like macromolecular aggregates composed of charged (macro)molecules. What we have specifically in mind are two rather dissimilar systems: on the one hand soft spheroidal charged lipid vesicles and on the other relatively hard icosahedral empty virus shells (capsids). In the former case the shell is composed of a quasi 2D fluid layer of lipid molecules \cite{Roux}, whereas in the latter the shell can be envisioned as a crystal-like assembly of tethered capsomeres \cite{Lorman}. Once assembled, such structures may become unstable if the amount of salt dissolved in the bathing solution, quantified by its ionic strength, is decreased, so that the screening of electrostatic repulsion becomes less efficient \cite{Butt}. For the macromolecular aggregates composed of complex molecules such as proteins, as in the case of virus shells, the amount of charge they carry may be modified by shifting the dissociation equilibrium via the solution pH \cite{Bo}. Apart from the changes in the ionic strength this presents yet another effective way to influence the stability of shell-like aggregates. In general, once the shell becomes unstable due to changes in the electrostatic interactions brought on by either the changes in ionic strength or pH, it may break up according to different scenarios.  We shall consider here only the conceptually simplest cases, and we shall show how the consideration of these simple cases leads to more general conclusions regarding shell stability. We shall apply our results both to fluid and tethered charged shells, so our findings should be of importance in the study of stability of charged fluid vesicles as well as virus capsids. 

The structure of the paper is as follows: we shall first introduce a simplified model of partially formed fluid and tethered spherical shells (Sec.~\ref{ssecA}). We shall then formulate the theory of electrostatic interactions for such simplified geometries based on the solution of either the complete Poisson-Boltzmann (PB) theory (Sec.~\ref{ssecB}) or its simplified derivative in the Debye-H\" uckel (DH) form (Sec.~\ref{ssecC}). The latter case will naturally lead us to introduce the concept of electrostatic line tension which we will show survives to a large extent in the complete Poisson-Boltzmann treatment (Secs.~\ref{ssecC} and~\ref{ssecD}). Based on this concept we will then describe the rupture of tethered as well as fluid shells and finally construct the appropriate stability ``phase diagram'' of the ruptured shells (Sec.~\ref{sectri}). We will compare our results with previous attempts at elucidation of the stability of spherical aggregates and conclude by pointing out the most important consequences of our theory (Secs.~\ref{secstiri} and~\ref{secpet}).

\section{PB level description of electrostatics of a partially formed charged spherical shell}

The PB theory of electrostatic interactions in ionic solutions is based on the mean-field approximation and is thus applicable only for sufficiently small charge densities on the surface of the macromolecules, low ion valencies, high medium dielectric constant, or high temperatures \cite{Wenbing}. In the case of monovalent salt solutions with low surface charge densities these limitations are not particularly severe and the results of the PB theory are quantitatively correct even when compared to more sophisticated approaches \cite{Hoda}. The limitations of the PB approach become practically important only in highly charged systems where ion-ion correlation effects begin to affect the electrostatic properties of the charged system \cite{Ali}. Though the accuracy of the PB approach can be systematically improved by perturbative corrections in the ionic fluctuations around the mean-field solution, eventually one has to give up the idea of a mean-field description altogether taking recourse in a fundamentally different reformulation of the electrostatic theory based on the concept of {\sl strong coupling} \cite{Wenbing}. We shall not delve any further into these reformulations assuming in what follows that the solution conditions we are describing are far removed from the limit of strong coupling. 

\subsection{Model of a partially formed spherical shell and modes of rupture of tethered and fluid shells}\label{ssecA}

\begin{figure}[ht]
\centerline{\epsfig {file=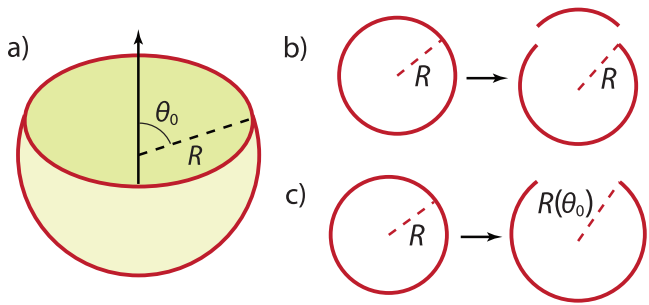,width=8.5cm}}
\caption{a) The geometry of a ruptured spherical shell. b) The mode of rupture of a tethered (solid) shell. 
c) The mode of rupture of a fluid shell. The amount of shell material in both modes of rupture is conserved in the process but is partitioned between the original shell and its progeny in the case of a tethered shell, or is contained completely in the progeny in the case of a fluid shell. }
\label{fig:fig1}
\end{figure}

The geometry of the partially formed spherical shell is shown in panel a) of Fig. \ref{fig:fig1}. The shape can be characterized by two parameters: the shell radius ($R$) and the opening angle ($\vartheta_0$). The surface of the shell is assumed to be homogeneously charged, so that the surface charge density $\sigma$ is a constant. Panels b) and c) of Fig. \ref{fig:fig1} sketch the ruptured shells in the case of tethered and fluid shells, respectively. 

We represent the rupture of a charged tethered shell in terms of a partition of the intact solid shell into two separate pieces described as two spherical caps with opening angles $\vartheta_0$ and $\pi - \vartheta_0$, with radii equal to that of the complete original shell ($R$). The surface charge density is assumed to remain the same in both pieces of the ruptured shell and equal to the one of the complete shell ($\sigma$). 

In the case of a fluid shell we assume that the rupture starts via formation of a pore that increases its radius as the rupture progresses and the pore opens up. During the poration the partially opened fluid shell is described as a spherical cap of radius $R(\vartheta_0)$, which is not equal to the radius of the original shell before the poration. We shall presume that the density of the shell material and the shell thickness remain the same during rupture and that the material of the fluid shell is conserved during poration. This also presumes that the total charge of the shell is conserved, so that the surface charge density does not change in the process of poration. 

\subsection{Electrostatic potential and free energy of a partially formed shell at PB level}\label{ssecB}

In the following, we shall assume that the bathing solution of the ruptured shell is an aqueous monovalent salt solution (relative permittivity $\varepsilon$) whose bulk concentration equals $c_0$. We describe the electrostatics of the partially formed shell in the PB framework \cite{Andelman2006,Grochowski2008}. Our approach to the problem is the same as that described in Ref. \onlinecite{SibRudi1}; however, the geometry is more involved in this case. In short, to obtain the electrostatic potential $\varphi$ in the solution that in this case depends on the radial coordinate, $r$, as well as the azimuthal angle, $\vartheta$, we solve the PB equation, 
\begin{equation}
\label{eq:pb}\nabla^2\varphi(r, \vartheta)=\frac{2e_0c_0}{\varepsilon\varepsilon_0}\sinh \Big( \beta e_0\varphi(r, \vartheta) \Big)\;.
\end{equation}
subjected to $\vartheta$-dependent standard electrostatic boundary condition on the shell surface of the form
\begin{equation}
\left.\frac{\partial\varphi(r, \vartheta)}{\partial r}\right|_{r=R^-}-\left.\frac{\partial\varphi(r, \vartheta)}{\partial r}\right|_{r=R^+}= \frac{\sigma}{\varepsilon\varepsilon_0}\quad\text{for}\quad\vartheta\in[\vartheta_0,\pi)\;,
\label{PBbc}
\end{equation}
where $\beta^{-1}=k_BT$ with $k_B$ the Boltzmann constant and $T$ temperature, $\varepsilon_0$ the vacuum permittivity, and $e_0$ the elementary charge. We assume an infinitely dilute solution of ruptured shells so that there are no interactions either between intact shells or between the products of the rupture.  We assume a salt bath in chemical equilibrium with the (ruptured) shell and its bathing ionic solution. The concentrations of positive and negative ions in the bath are equal, $c_+^0=c_-^0=c_0$\,. 

Once the electrostatic potential is obtained, the free energy of the system can be constructed by standard methods as detailed in Ref. \onlinecite{SibRudi1}. The electrostatic part of the (Helmholtz) free energy ($F_{el}$) can be evaluated in various equivalent ways but is most conveniently calculated from the charging process
\begin{equation}
F_{el} = \oint \mathrm{d}S \int_0^{\sigma} \varphi(r = R, \vartheta)\, \mathrm{d}\sigma\;,
\label{freen}
\end{equation}
where the upper bound of the integration over the surface charge density is set by the value of the surface charge residing on the surface $S$ \cite{Verwey}. The above form of the free energy can be easily simplified in the case of the DH linearization of the PB theory by acknowledging the fact that the DH electrostatic potential depends linearly on the charge density.

\subsection{Analytical solution of the problem in DH approximation}\label{ssecC}

When the electrostatic potentials in the electrolyte solution are small, $\beta e_0 \varphi \ll 1$, the PB equation can be linearized leading to the DH equation of the Helmholtz type whose solutions are well known. 

\subsubsection{DH equation, boundary condition, and free energy}

Linearizing the PB equation [Eq. (\ref{eq:pb})], the DH equation is then obtained as
\begin{equation}
\nabla^2 \varphi (r, \vartheta) - \kappa^2 \varphi (r, \vartheta) = 0\;,
\label{eq:helm}
\end{equation}
where $\kappa$ is the inverse DH screening length given by 
\begin{equation}
\kappa^{-1} \equiv \sqrt{\varepsilon\varepsilon_0/2 \beta e_0^2 c_0}\;.
\end{equation}

The electrostatic problem of a shell with radius $R$ that has a hole characterized by opening angle $\vartheta_0$ can be posed also in slightly different terms -- we may treat it as a problem of a {\sl complete} spherical shell but with an azimuthally varying surface charge density, $\sigma(\vartheta)$. In this case the surface charge density assumes a Heaviside-like form in terms of the azimuthal angle 
\begin{equation}
\sigma(\vartheta)=\left\{\begin{array}{rl}
0 & \text{, }\; \vartheta < \vartheta_0 \\
\sigma & \text{, }\; \vartheta_0 < \vartheta < \pi
\end{array}\right.\;.
\label{eq:bndry-not}
\end{equation}
The boundary condition in Eq. (\ref{PBbc}) then becomes
\begin{equation}
\left. \frac{\partial \varphi (r, \vartheta)}{\partial r} \right|_{R=R_{-}} - \left. \frac{\partial \varphi (r, \vartheta)}{\partial r} \right|_{R=R_{+}} = \frac{\sigma(\vartheta)}{\varepsilon \varepsilon_0}\;.
\label{eq:bndry}
\end{equation}

The potential that is a solution of Eq. (\ref{eq:helm}) can be written as a superposition \cite{Abram}, 
\begin{equation}
\varphi (r, \vartheta) = \sum_{l=0}^{\infty} \frac{1}{\sqrt{r}} \left[ a_l I_{l+\frac{1}{2}} (\kappa r) + b_l K_{l+\frac{1}{2}} (\kappa r) \right] P_l (\cos \vartheta)\;,
\end{equation}
where $a_l$ and $b_l$ are the expansion coefficients, $P_l$ are Legendre polynomials, and $I_{l+\frac{1}{2}}$ and $K_{l+\frac{1}{2}}$ are the modified spherical Bessel functions of the first and the third kind, respectively.

To obtain the free energy of the shell corresponding to the appropriate solution of the DH equation (see Appendix \ref{sec:soldh}), one needs to evaluate the charging integral given by Eq. (\ref{freen}). 
The electrostatic free energy in the DH approximation can then be obtained in the form
\begin{equation}
F_{el}^{DH} = \pi R^2 \sigma \int_{\vartheta_0} ^{\pi} \mathrm{d} \vartheta \sin \vartheta \,\varphi(R, \vartheta)\;, 
\label{neweqfree}
\end{equation}
yielding finally 
\begin{equation}
F_{el}^{DH} = \frac{\pi  R^2 \sigma ^2}{ \varepsilon\varepsilon_0 \kappa}\,\frac{1}{2}\sum_{l=0}^{\infty} 
\frac{{\cal F}_3^2 (l, \vartheta_0)}{(2l+1){\cal F}_0 (l,\kappa R)}\;.
\label{eq:finalenerg}
\end{equation}
The functions ${\cal F}_3 (l, \vartheta_0)$ and ${\cal F}_0 (l,\kappa R)$ are defined in Eqs. (\ref{firstdef}) and (\ref{seconddef}).

\subsubsection{Asymptotic expansion and limiting form of free energy}

The numerical evaluation of Eq. (\ref{eq:finalenerg}) does not pose significant problems, but the analytical limits that may be of interest turn out to be more difficult to come by. These were obtained using routines from Mathematica \cite{Wolfram}. 

One is first interested in the limit when $\vartheta_0 \rightarrow 0$. This represents the complete shell without a hole. The result is
\begin{equation}
\lim_{\vartheta_0 \rightarrow 0} F_{el}^{DH} = \frac{\pi  R^2\sigma^2}{ \varepsilon\varepsilon_0 \kappa}\frac{2}{1+\coth\kappa R}\;,
\label{eq:DHlimitclosed}
\end{equation}
which is the same as Eq. (15) in Ref. \onlinecite{SibRudi1}.

Some insight can be obtained by analyzing the asymptotic forms of the Bessel functions featuring in the free energy. By using the asymptotic expansions, given in Refs.~\onlinecite{Watson} and~\onlinecite{Arfken}, we obtain
\begin{equation}
F_{el}^{DH} \asymp \frac{\pi  R^2 \sigma ^2}{ \varepsilon\varepsilon_0 \kappa}\,\frac{1}{2}\sum_{l=0}^{\infty}\frac{{\cal F}_3^2(l,\vartheta_0)}{2l+1}\frac{1}{2}\left[1-\frac{l(l+1)}{2\rho^2}\right]+{\cal O}\left(\frac{1}{\rho^3}\right)\;,
\label{eq:asymptDH1}
\end{equation}
where we have introduced $\rho \equiv \kappa R$. It turns out that the asymptotic expansion is problematic for any finite value of $\rho$, so that the free energy for an open shell can only be obtained by a significantly extended summation of the {\sl nonasymptotic} $l$ series (detailed discussion is given in Appendix~\ref{sec:appz}). Since effectively no asymptotic result can be derived via a simplistic expansion of the free energy in Eq. (\ref{eq:finalenerg}), one needs to approach the problem by a different route. Note here, however, that the zeroth-order terms (proportional to $\rho^0$) of the expansion [Eq. (\ref{eq:asymptDH1})] may well be valid, as they do not depend on the effects of the edge and its screening. We shall establish in what follows that this is indeed the case.

Problems with the range of validity of the asymptotic form of the electrostatic free energy obtained directly from the solution of the DH equation lead us to an alternative formulation that would avoid the pitfalls of the infinite summation due to the sharp edges of the partially opened shell. It is based on the simple equivalence between two ways of calculating the electrostatic energy of the system: it can be thought of as a result of the charging of the system or as a result of the direct interactions in the system. Both viewpoints are equivalent \cite{Verwey}.

By taking now the second approach and summing the interactions over the surface of the incomplete shell (see Eq. (\ref{eq:DHdiscrete}) and the derivation in the Appendix~\ref{sec:appa}), one obtains the form of the free energy in the limit when $\kappa R \gg 1$ but remains finite:
\begin{equation}
\lim_{\kappa R \gg 1} F_{el}^{DH} = \frac{\pi R^2 \sigma^2 }{ \varepsilon\varepsilon_0 \kappa}\left[\frac{1+\cos\vartheta_0}{2}-\frac{f}{\kappa R}\sin\vartheta_0\right]\;,
\label{eq:DHapprox}
\end{equation}
where $f$ is a numerical constant, $f \sim 0.12$. The second term in Eq. (\ref{eq:DHapprox}) arises from the fact that the charges on the shell close to the open shell edge lack some of the neighboring charges with respect to the charges away from the edge. The parameter that determines the effective distance of a particular charge from the shell edge is the screening length, $1 / \kappa$. Since the length of the open boundary of the shell is
\begin{equation}
{\cal L} = 2 \pi R \sin \vartheta_0\;,
\end{equation}
the second term is obviously proportional to ${\cal L}$ and can be interpreted as the {\sl interaction renormalized line tension} energy of the edge. The interaction renormalization of mesoscopic material properties, line tension in this case, is a ubiquitous feature of molecular interactions in matter \cite{Landau} and its appearance should also not come as too big a surprise in the present context.  

Note, however, that since the result was derived in the limit when $\kappa R \gg 1$ but still finite, the second term is significantly smaller from the first one as the contribution of the edge to the total electrostatic energy of the shell is small in this limit. For e.g. viruses, the condition $\kappa R \gg 1$ is typically well obeyed in physiological conditions \cite{Kegel1,SibRudi1}. 

\begin{figure}[!ht]
\centerline{
\epsfig {file=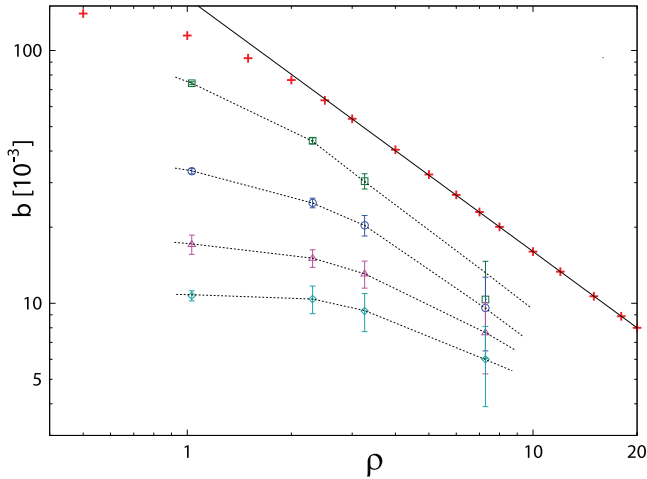,width=8.5cm}
}
\caption{The results of the fit of Eq. (\ref{eq:fitDH}) to the DH solution (pluses). The full line shows the $b_0/\rho$ behavior with $b_0\approx0.161$\,. The empty symbols show the behavior of the fit to Eq. (\ref{eq:fitDH}) of the full PB solution calculated for $R=10.04$ nm, for different surface charge densities: $\sigma=$ 0.1 $e_0/\text{nm}^2$ (squares), 0.4 $e_0/\text{nm}^2$ (circles), 0.8 $e_0/\text{nm}^2$ (triangles), and 1.2 $e_0/\text{nm}^2$ (diamonds). Dashed lines are guides to the eye. The numerical solution of the PB solution become less reliable with increasing $\rho$\, as the errorbars indicate. A noticeable effect in PB case is the dependence of the line tension coefficient $b$ on the surface charge density.
}
\label{fig:linefit}
\end{figure}

Now that we have found an appropriate form of the electrostatic free energy in the limit $\infty > \kappa R \gg 1$ we can check how well the {\sl exact} (non-asymptotic) DH solution [Eq. \ref{eq:finalenerg}] fits to it. One should recall that we were not able to obtain this form of the free energy by merely analyzing the asymptotic form of the DH free energy. In accordance with Eq. (\ref{eq:DHapprox}), we take the fitting form 
\begin{equation}
\label{eq:fitDH}F_{fit}(\vartheta_0;a,b)=a\left(\frac{1+\cos\vartheta_0}{2}-b\sin\vartheta_0\right)
\end{equation}
where $a$ and $b$ are numerical constants to be determined by the fit to Eq. (\ref{eq:finalenerg}). The fit is shown in Fig.~\ref{fig:linefit}. 
We find a non-zero value for $b$ of the form
\begin{equation}
\label{eq:bcoef}b=\frac{b_0}{\kappa R}\quad\text{where}\quad b_0\approx0.161\quad.
\end{equation}
This is completely consistent with the approximate form in Eq. (\ref{eq:DHapprox}), and the value of $f$ that we obtained from the approximate analysis in $\kappa R \gg 1$ regime 
is not far off from the value of $b_0$.

The deviations from the $1/\rho$ behavior are apparent in the regime where $\rho\leq1$\,, where in fact we do not observe the part with $\sin\vartheta_0$ dependence predicted by Eq. (\ref{eq:fitDH}) and the line tension renormalization concept is of limited use there. In this regime, the screening from the inside of the shell is incomplete since the volume of the shell interior is too small to contain all the salt ions required to screen the shell charge, i.e. $R$ is too small since $\rho\sim1$\,. Thus, the line tension renormalization is different from the form implied by Eq. (\ref{eq:DHapprox}). This effect only becomes important for small angles $\vartheta_0$ when there is a clear separation between interior and exterior. For larger values of $\vartheta_0$\,, when the shell looks more like a disk than a sphere, the separation of space into interior and exterior does not make sense and $\sin\vartheta_0$ behavior holds for large enough angles, even when $\rho \sim 1$\,.

\subsection{Solution of nonlinear PB equation}\label{ssecD}

We now solve the full nonlinear PB equation [Eq. (\ref{eq:pb})] and evaluate the corresponding electrostatic free energy [Eq. (\ref{freen})].

Due to the axial symmetry, the problem is effectively two-dimensional, and the solution can be sought in the domain $(r,\vartheta)\in[0,\infty)\times[0,\pi)$. To solve Eq.~(\ref{eq:pb}), we first discretize the domain and the PB equation for the electrostatic potential using finite differences. Typically, we used about 1200 points in the radial and 500 in the polar direction in the nonlinear regime of parameters. The largest radial point has to be chosen such that the potential at this point is very nearly close to zero. The band matrix system obtained in this way is solved using a damped nonlinear Newton method~\cite{NRbook}, iterated until desired convergence is reached. 

\begin{figure}[ht]
\centerline{
\epsfig {file=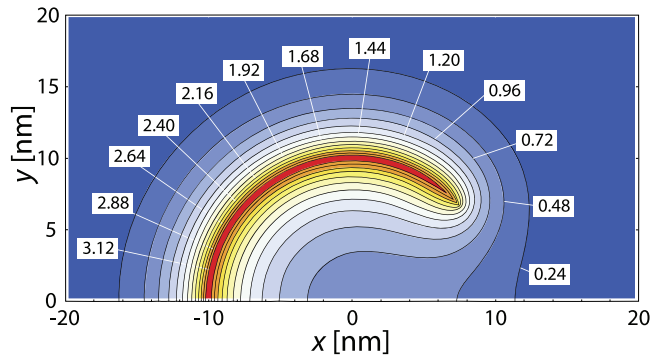,width=8.5cm}
}
\caption{The solution of the PB equation for the dimensionless electrostatic potential $\beta e_0\varphi$\,. The shell radius is $R=10.04$ nm, and the opening angle $\vartheta_0=0.235\pi$\,. The shell surface charge density is $\sigma=0.4$ $e_0/\text{nm}^2$\,, and the bulk concentration of salt ions is $c_0=10$ mM.
}
\label{fig:potential}
\end{figure}

Figure~\ref{fig:potential} shows the numerical solution of Eq.~(\ref{eq:pb}) for the dimensionless electrostatic potential $\beta e_0\varphi$.

\begin{figure}[ht]
\centerline{
\epsfig {file=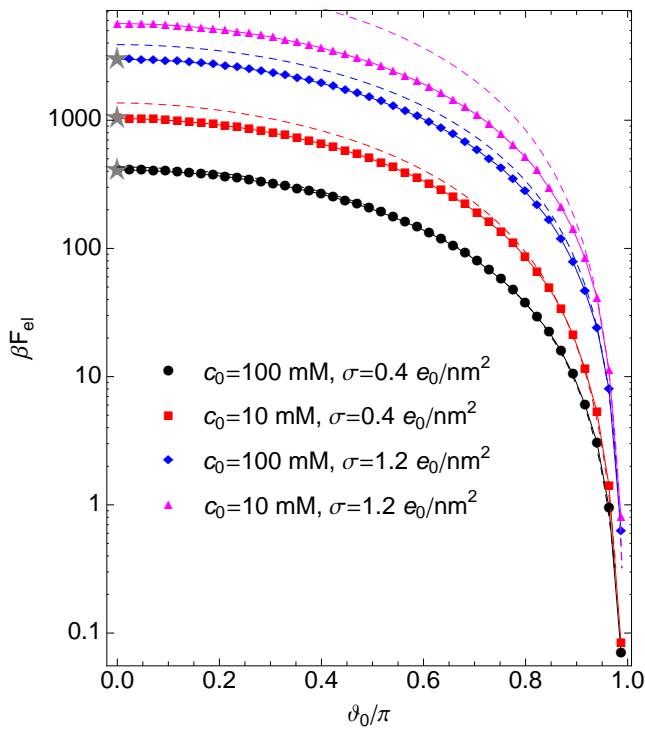,width=8.5cm}
}
\caption{Electrostatic free energy of the system as a function of the opening angle, for a capsid of radius $R=10.04$ nm and several values of surface charge densities and salt concentrations, as denoted in the body of the figure. The dashed lines show the solution obtained in the DH approximation; the black dashed line aligns almost completely with the numerical PB solution. The star symbols show the PB results for a closed capsid ($\vartheta_0=0$) taken from Ref.~\onlinecite{SibRudi1}.
}
\label{fig:FEtheta}
\end{figure}

Figure~\ref{fig:FEtheta} shows the dependence of the electrostatic free energy of the system on the opening angle of the shell. In the limit when $\vartheta_0\to0$\,, the problem reduces to the one of a closed shell studied in Ref.~\onlinecite{SibRudi1}; these results are also shown in Fig.~\ref{fig:FEtheta}. The observed agreement nicely parallels the one obtained in the DH limit as well, since we have previously demonstrated that in the limit when $\vartheta_0\to0$ the analytical DH result [Eq. (\ref{eq:finalenerg})] reduces to the form previously derived in Ref. \onlinecite{SibRudi1}. We also compare the DH results obtained from Eq. (\ref{eq:finalenerg}) and shown in Fig. \ref{fig:FEtheta} with the numerically exact PB results.

\begin{figure}[ht]
\centerline{
\epsfig {file=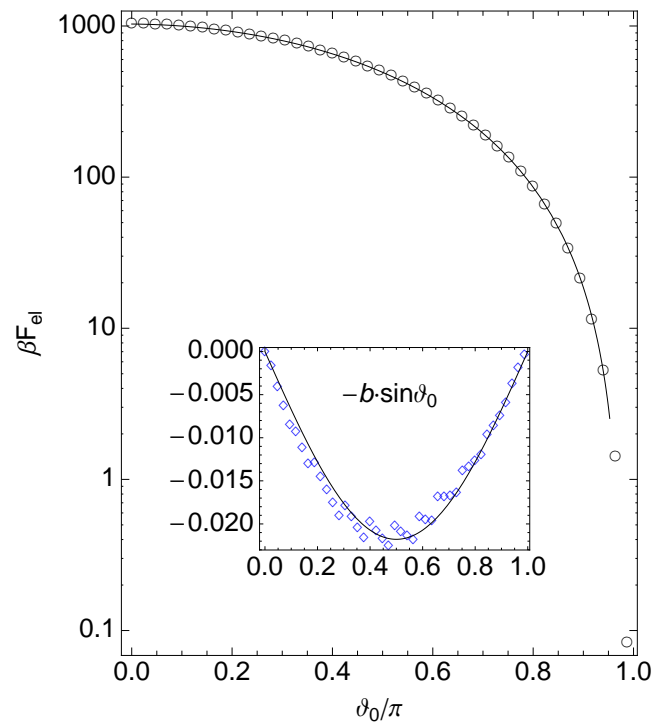,width=8.5cm}
}
\caption{Fit of the angular dependence of the solution of the PB theory to the form $F_{fit}(\vartheta_0)=a[(1+\cos\vartheta_0)/2-b\sin\vartheta_0]$, for a shell of radius $R=10.04$ nm and surface charge density $\sigma=0.4$ $e_0/\text{nm}^2$\,. Salt concentration is $10$ mM. The inset shows a separate fit of $-b\sin\vartheta_0$ to the numerical solution, from which the cosine dependence was removed, i.e. $F_{num}(\vartheta_0)/F_{num}(0)-(1+\cos\vartheta_0)/2$\,. Note that in the latter case, a slight effect of numerical error can be discerned.
}
\label{fig:FEfit}
\end{figure}

We now examine the concept of electrostatic renormalization of the line tension that we introduced before at the DH level. The question is whether and how the concept survives in the numerically exact solution of the PB theory  in the region of parameters where the DH approximation does not apply. 
To test the usefulness of the line tension concept even outside the DH limit, we have compared the angular dependence of the electrostatic free energy obtained from a numerical solution of the complete PB equation to the form of Eq.~(\ref{eq:fitDH}). The result of a numerical solution of the PB equation is shown in Fig.~\ref{fig:FEfit}, and the values of line tension coefficients $b$ obtained from fits to Eq.~(\ref{eq:fitDH}) are summarized in Fig.~\ref{fig:linefit}.

It is obvious from Fig.~\ref{fig:linefit} that the line tension dependence describes the behavior of the electrostatic free energy quite well even in the case of complete nonlinear PB theory. The numerical data, even in the case of very small salt concentrations (e.g. 1 mM) where the DH approximation does not apply, can be fitted to the general form of Eq.~(\ref{eq:fitDH}), although the electrostatic contribution to the line tension coefficient, $b$, may be quite different from the DH prediction. As expected, the numerical data approach the DH prediction as the salt concentration increases, as is usual since the electrostatic potentials in the solution decrease \cite{SibRudi1}. Contrary to the prediction of the DH theory [Eq. (\ref{eq:bcoef})] the line tension coefficient shows dependence on the surface charge density (see Fig.~\ref{fig:linefit}), particularly in regimes where the DH limit does not hold. 

\section{Rupture of shells}\label{sectri}

Regardless of the smallness of the electrostatic contribution to the edge tension energy, particularly when $\rho \gg 1$, it is precisely this term that induces the shell rupture. One notices that electrostatics favors the opening of the edge. It is the opposing attractive interactions of the vdW/hydrophobic type that keep the shell together. We shall assume that the attractive interactions are of short range, so that the attractive part of the cohesive energy can be approximately associated with ``bonds'' between the molecular constituents of the shell. Hydrophobic and van der Waals interactions could be conceived to be of such nature. These soft molecular bonds are broken upon rupture, so that the part of the cohesive energy that is lost in the process is proportional to the edge length. 

This cohesive energy change is different for the case of tethered as opposed to fluid shells. In the former case, {\sl two} partially opened shells are created upon rupture, whereas in the latter case the rupture creates a {\sl single} partially opened shell. Also in the case of the fluid shell rupture one has to properly include the change in the curvature energy upon rupture, since the pore changes the radius of curvature of the shell (see Fig. \ref{fig:fig1}). We thus treat the cases of tethered and fluid shells separately.

\subsection{Rupture of tethered shells}

Tethered shells with a spherical geometry show a more or less pronounced faceting, but in what follows we will confine ourselves to the ideal sphere, ignoring the effects of (icosahedral) faceting \cite{shellshapes1,shellshapes2} for the sake of calculational feasibility.

We denote the line tension parameter arising solely from the attractive interactions by $\gamma$, so that the total energy required to form an edge of length ${\cal L}$ (or to separate a complete shell into two spherical caps with opening angles $\vartheta_0$ and $\pi - \vartheta_0$) is 
\begin{equation}
F_{edge} = \left( \gamma - \frac{b_0 \sigma ^2}{ \varepsilon\varepsilon_0 \kappa^2} \right) {\cal L}\;.
\label{eq:edge_tether}
\end{equation}
Note here that $F_{edge}$ contains the electrostatic contribution to the edge energy of the incomplete shell twice, because upon rupture, two incomplete shells are formed, both having an open edge (note also that, since both spherical caps have the same curvature, we do not need to include the change in the bending energy here).

\begin{figure}[ht]
\centerline{
\epsfig {file=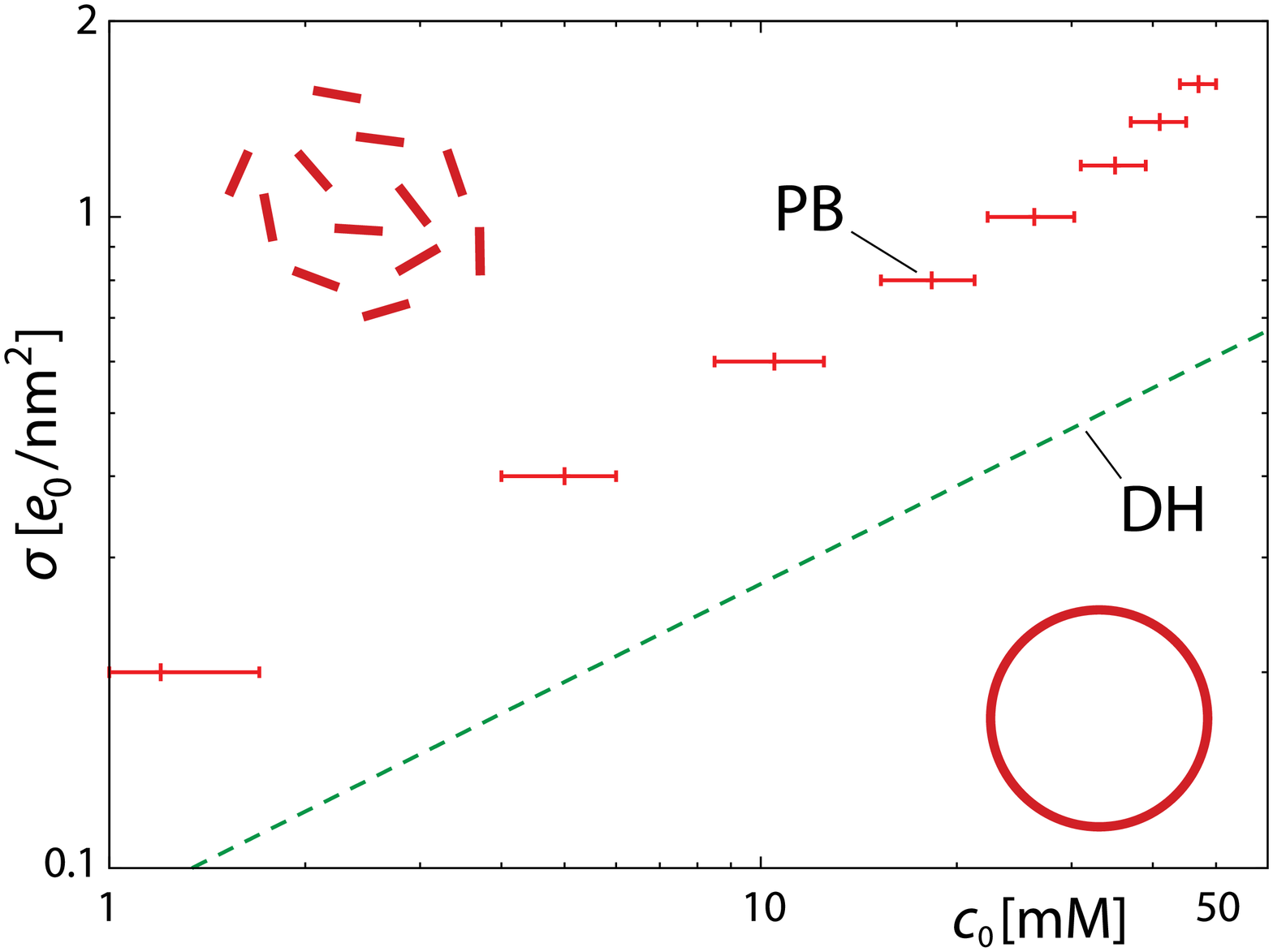,width=8.5cm}
}
\caption{Symbols: Instability line of tethered shells towards rupture obtained by numerical PB calculation with $\gamma = 1$ $k_B T$/nm and shell radius $R=$10.04 nm. Dashed line: Debye-H\"{u}ckel approximation, $\sigma = e_0 \sqrt{2 \beta \gamma c_0 / b_0}$. In PB case, the points $(\sigma, c_0)$ of zero edge energy were obtained by interpolation of numerical results for several surface charge densities $\sigma$ and salt concentrations $c_0$, hence the errorbars.
} 
\label{fig:fig_DH_stabil}
\end{figure}

When the quantity in the brackets (the edge tension parameter renormalized by electrostatic interaction) in the rhs of Eq. (\ref{eq:edge_tether}) becomes zero, the rupture spontaneously takes place. Since the energy gained is proportional to the edge length, one can expect the breakage of all the bonds, a complete disintegration of the tethered shell into its (macro)molecular constituents. In the context of our model, a particular macromolecule may be imagined as a small spherical cap with a certain edge length. The rupture of a sphere would thus correspond to disintegration of {\sl all} the small spherical caps from the assembled structure. Such final state would have the lowest free energy of all others that could be formed by assembling the small spherical caps (macromolecules) in several larger cap-like aggregates (ruptured parts of the shell); this conclusion holds as long the concept of line tension makes sense. This finding enables us to draw a diagram of the tethered shell stability, i.e. to draw a line separating the stable shells from the ruptured ones in the $\sigma - c_0$ plane (with $\gamma$ assumed to be constant and independent of $\sigma$ and $c_0$). This phase diagram is shown in Fig. \ref{fig:fig_DH_stabil}.

There arises a question on the reliability of the instability line obtained in this way. Namely, the DH approximation is reliable only when the electrostatic potential $\varphi$ in the solution is small, i.e. when $\beta e_0 \varphi \ll 1$. We have shown, however, that the concept of the electrostatic renormalization of the line tension carries over from the DH limit onto the full PB theory almost unchanged. The most important difference between the DH and the complete PB theory is that the electrostatic renormalization of the line tension is an effect significantly smaller than predicted by the DH approximation (see Fig.~\ref{fig:linefit}). This is also the reason that in the  $(c_0,\sigma)$\, phase diagram the shells are stable in a larger regime than predicted by the DH limit (see Fig. \ref{fig:fig_DH_stabil}). 

\subsection{Poration (rupture) of fluid membrane shells}

As already discussed, fluid membrane shell can rupture without loss of material by simply rearranging the constituent (lipid) molecules so that the hole opens up. It is known that the vesicles may form holes in spite of the line tension (exposure of hydrophobic parts of the constituent molecules) if their surface is charged~\cite{Betterton1999}. Experimentally, such pores have been observed for instance in the so-called red blood cell ghosts at low salt concentrations~\cite{Lieber1982}. %

Here we briefly investigate the possibility that the charge on the vesicle induces a different kind of surface deformation. Instead of vesicle poration, one may think of a situation where the vesicle separates into two smaller, completely closed vesicles, so that the edge does not form in the process. This scenario is similar to the Rayleigh instability of charged droplets \cite{Rayleigh} discussed in the Introduction. However, since the electrostatic energy of the charged shell in the DH regime (and when $\kappa R \gg 1$) is proportional to \cite{SibRudi1} $\sigma^2 R^2$, and in the process of transformation into two smaller closed vesicles the area remains the same, there is no gain in the electrostatic energy in the final state of the system. This type of instability is of importance only in the poor screening, near-Coulomb regime, where the electrostatic energy of the vesicle is proportional to the third power of the vesicle radius \cite{SibRudi1}. We shall neglect these effects in the following and concentrate exclusively on the charge induced poration.

It can be envisaged that the vesicle with a hole may have quite complex geometry \cite{vesiclephasediagram}, but we shall approximate it by a sphere with a missing cap as in the previous sections. Due to the way the pore forms we consider the sequence of shapes that have the same number of molecules and thus the same area, but different radii, or equivalently, the opening angles of the pore. Such a sequence would thus start from a fully closed vesicle and open to a flattened-out disk. We parametrize this sequence of shapes with an opening angle $\vartheta_0$ of the formed pore.

For the total area of the vesicle with a pore to equal the area of a vesicle without pore the radii of the shapes in the sequence must grow with the opening of the pore, $R=R(\vartheta_0)$\,. Given a closed vesicle with radius $R_0$ and area $A=4\pi R_0^2$\,, some elementary geometry then gives
\begin{equation}
R(\vartheta_0)=R_0\sqrt{\frac{2}{1+\cos\vartheta_0}}\;,
\end{equation}
and similarly for the $\rho(\vartheta_0) = \kappa R(\vartheta_0)$\,, where we now denote $\rho_0=\kappa R_0$\,.

We need to consider the influence of the elastic bending energy of the vesicle as well, since the curvature of the shell changes as the pore opens. In the classical curvature model, the local elastic bending energy is in the lowest order composed of two terms containing mean and Gaussian curvature~\cite{Seifert1997}. With the two curvatures we associate two elastic constants, $\kappa$ and $\kappa_G$, called bending rigidity and Gaussian bending rigidity, respectively. As we are interested only in a piecewise spherical surface, both radii of curvature are the same.
Using $R(\vartheta_0)$ defined above we see that the curvature free energy can be written as \cite{SiberMaj}
\begin{equation}
F_{bend}=2\pi\mathcal{K}\,\frac{1+\cos\vartheta_0}{2}\;.
\end{equation}
Here the Gaussian and the bending elastic moduli have been combined into an {\sl effective bending rigidity parameter} of the vesicle, $\mathcal{K}=4(\kappa+\kappa_G/2)$\,. In the equation above we have assumed that the spontaneous curvature of the shell is zero, although the case with non-vanishing value of the spontaneous curvature can be easily studied along the lines that we present in the following.

We now derive the complete free energy $F_{total} = \gamma {\cal L} + F_{bend} + F_{el}^{DH}$ including thus the non-electrostatic line tension energy, the bending energy, and the electrostatic energy. The latter we derive first within the DH approximation confining ourselves to the limit $\kappa R(\vartheta_0) \gg 1$. In this case the electrostatic part of the free energy can be derived in the DH limit by inserting $R(\vartheta_0)$ into Eq. (\ref{eq:DHapprox}) and obtaining for the total free energy
\begin{equation}
F_{total} = \frac{\pi \sigma^2 R_0^2}{ \varepsilon\varepsilon_0 \kappa} + F_{total}(\vartheta_0)\;,
\end{equation}
where
\begin{eqnarray}
\frac{F_{total}(\vartheta_0)}{2\pi{\cal K}} &-& \frac{1+\cos\vartheta_0}{2} = \nonumber\\
&=& V_F\left(1-\frac{\gamma_{el}}{\gamma}\right)\sqrt{2(1-\cos\vartheta_0)}\;.
\end{eqnarray}
Here we introduced the vesiculation index $V_F$ and the electrostatic line tension $\gamma_{el}$ as
\begin{equation}
V_F=\frac{R_0\gamma}{{\cal K}}=\frac{{\cal L}_0\gamma}{2\pi{\cal K}} \qquad {\rm and} \qquad \gamma_{el}=\frac{b_0\sigma^2}{2\varepsilon\varepsilon_0\kappa^2}\;,
\end{equation}
with ${\cal L}_0 = 2 \pi R_0$. Note that $V_F$ is defined as {\sl half} the value introduced in Ref.~\onlinecite{Fromherz1983}. The only parameters determining relative contributions of different free energies are thus $V_F$\,, the ratio $\gamma_{el}/\gamma$\,, and $\rho_0$\,.

The specific rewrite of the energetics of the shell in the DH limit with $\rho\gg1$ reduces the problem to the one studied by Fromherz in Ref.~\onlinecite{Fromherz1983}, where only the line tension and curvature contributions were considered. In our case, however, we have an additional possibility of the line tension energy being both positive and negative as a consequence of the electrostatic effects. Without electrostatic interactions it is always costly to form edges due to residual attractive interactions giving rise to non-electrostatic line tension energy, $\gamma$. With electrostatic interactions included, it may be favorable to form holes. This happens when $(1-\gamma_{el}/\gamma)\leq0$\,. For pores in the vesicle to open up spontaneously, the free energy has to fall off as a function of the opening angle. 

\begin{figure}[!htp]
\centerline{\epsfig {file=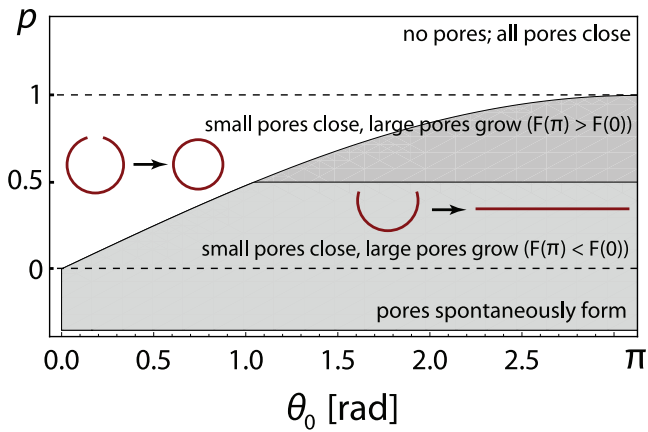,width=8.5cm}}
\caption{The ``phase diagram'' of the ruptured capsids that shows the sign of $\partial F_{total} / 
\partial {\vartheta_0}$ (gray regions). In light-gray (dark-gray) regions, the fully opened, disk state of the membrane has lower (higher) free energy than the spherical state. 
In the white regions, the free energy always increases with $\vartheta_0$ and the vesicle is stable towards pore formation.}
\label{fig:fig2}
\end{figure}

It is convenient to introduce the ratio of energy parameters, 
\begin{equation}
p=V_F(1-\gamma_{el}/\gamma)\;.
\label{eq:our_p}
\end{equation}
The total shell free energy can be written as a functional whose universal behavior depends on the parameters $p$ and $\vartheta_0$. In Fig.~\ref{fig:fig2}, we plot the function
\begin{equation}
\frac{1}{2\pi{\cal K}}\frac{\partial F_{total}}{\partial\vartheta_0} = \frac{\sin \vartheta_0}{2} \left( \frac{p}{\sin \frac{\vartheta_0}{2}} - 1 \right)\;,
\end{equation}
which allows us to separate regions where it is larger or smaller than zero (note that ${\cal K}>0$). The ``phase diagram'' shows the type of information already studied by Fromherz, but modified by electrostatic effects in our case. One can see three different regimes: The first one takes place for when $p<0$, where even pores of the smallest size are unstable and they thus spontaneously form. For $0<p<1$\,, only the pores of certain critical radius will spontaneously grow, otherwise they will close. The critical pore opening angle (i.e. the critical pore radius) for given $p$ is at the point where the white region meets the gray regions. For $p>1$, there are no stable pores. The second regime, where $0<p<1$, can be divided in two parts: in the first one ($0<p<0.5$) the disk state has lower free energy than the sphere state, and in the second one ($0.5<p<1$) it is the other way around; both regimes are indicated in Fig. \ref{fig:fig2}.

The PB approach only shifts the borders in the DH phase diagram, which remains qualitatively the same, i.e. we observe no additional effects, metastable pores in particular. This is a consequence of the validity of the line tension concept even in the PB limit. As we have shown (see Fig. \ref{fig:linefit}), the PB solution only predicts a lower value of line tension, the more so the larger the surface charge density. This effect cannot change the salient features of the pore energetics discussed previously at the DH level.

We note in passing that the formation of pores could be also described as {\sl pore nucleation}, emphasizing in this way the similarity of the process to the physics of the classical theory of nucleation \cite{nucleationreference}. As is the case there, we also have a critical size of the pore required in order to start the pore nucleation process, which extends the pore to the maximal possible size, leaving the fluid shell in the disk state.

Let us now relate the results obtained for fluid shells to those previously obtained for the tethered shells. In the case of tethered shells, the energy contribution of bending rigidity of the shell was of no importance for the rupture mode. If we put $\mathcal{K} = 0$, then the phase diagram shown in Fig. \ref{fig:fig2} separates in two regions only, the one where $1-\gamma_{el}/\gamma>0, \; p \rightarrow +\infty$, and the shell is stable towards pore formation, and the other where $1-\gamma_{el}/\gamma<0, \; p \rightarrow -\infty$, and pores spontaneously form. Without consideration of the bending rigidity, the region of parameters where the pores of critical radius grow and smaller ones close does not exist. The stability line is given by the condition $1-\gamma_{el}/\gamma = 0$ which gives 
\begin{equation}
\gamma - \frac{b_0}{2}\frac{\sigma^2}{\varepsilon \varepsilon_0 \kappa^2} = \gamma - \gamma_{el} = 0\;.
\end{equation}
This is the same critical relation as the one following from Eq. (\ref{eq:edge_tether}) up to the multiplicative constant of 2 which arises due to two electrostatically renormalized open edges that exist in the case of tethered shells. Therefore, when $\mathcal{K}=0$, the discussion regarding the stability of fluid shells towards poration is quite similar as in the case of tethered shells discussed previously. As in that case, in a region of charge densities and salt concentrations above the full line in Fig. \ref{fig:fig_DH_stabil} ($\sigma \propto \sqrt{c_0}$; large $\sigma$'s, small $c_0$'s), the shells will spontaneously rupture.

\section{Application of results to viruses and vesicles}\label{secstiri}

An infinitely thin, spherical, and uniformly charged tethered shell may seem like an oversimplified model of an icosahedral virus capsid. Indeed, icosahedral viruses are not perfectly spherical \cite{shellshapes1,shellshapes2}, their capsids are several nanometers thick, and the charges on their constituent proteins are not uniformly distributed on the capsid surface but show variations consistent with their icosa(delta)hedral symmetry \cite{Konecny}. One could formulate a more complicated electrostatic model of a virus capsid, as has been done in the literature \cite{SibRudi1,Konecny}. One should remember, however, that our primary interest is in the capsid rupture and the energetics of edge formation. The almost analytical type of insight in this matter that we detailed in the previous sections would be difficult to obtain with more complex capsid models that would need to rely completely on numerical methods \cite{Konecny}.

The assumption of vanishing thickness of the shell may appear as the most worrying feature of our model when applied to real viruses. 
Indeed, typical capsid thickness and the Debye-H\"{u}ckel screening length at physiological ionic strength are quite comparable. As the charge density 
in general varies across the capsid thickness \cite{Anzeinpreparation}, one may infer that this could invalidate the application of our model to real viruses. This is not the case since the protein material of the capsid cannot be treated as totally permeable to salt ions, so that there is no screening of charge across the capsid thickness. Instead, the protein material should be treated as a medium with small relative dielectric constant, as has been done e.g. in Refs. \onlinecite{SibRudi1} and \onlinecite{Prinsen1}. This has been studied in some detail previously in Ref \onlinecite{SibRudi1}, where it was demonstrated that the most important features of the spherical shell energetics are already contained in the model of an infinitely thin charged shell. This also includes the analytical behavior of the electrostatic energy, i.e. its scaling with the shell radius and the surface charge density. 

Our calculations predict an instability of a charged tethered shell that depends on its surface charge density and the concentration of 
monovalent salt in the solution into which it is immersed. The tethered shells become unstable either with (i) increase of the surface charge density, or (ii) with decrease of the salt concentration. This holds both in the DH and PB approaches to the problem as can be seen in Fig. \ref{fig:fig_DH_stabil}. These results should apply to viruses, but to the best of our knowledge, there are no systematic studies of empty capsid stability with respect to changes in pH and salinity. There are studies of the conditions for the self-assembly of empty capsids \cite{butler,lavelle,wing_hepB,zlotnick_hepB}. These are not directly applicable to the problem of interest to us, since the assembly necessarily involves change in the entropy of dissolved proteins (monomers) \cite{Zandi,SiberMaj}. Assuming that the capsid disassembly and assembly are two faces of the same thermodynamical {\sl equilibrium} process, one finds that the capsid stability depends on the critical concentration of proteins which is related to the free energy gain (loss) upon assembly (disassembly) \cite{Zandi,SiberMaj}. Thus, in sufficiently low concentrations, the capsids should be unstable, regardless of the strength of the protein contacts in the capsid. However, it has been demonstrated that the disassembly of hepatitis B virus capsids proceeds exceedingly slowly and that the capsids are stable for days in the low-concentration conditions where the disassembly should take place \cite{Zlotnick_hysteresis}. This has led the authors of Ref. \onlinecite{Zlotnick_hysteresis} to conclude that the capsid disassembly and assembly reactions are not in the equilibrium. This is why our approach should be of use, as it concentrates on the {\sl enthalpic} contribution to the shell energetics, and the disassembly induced in such a way should depend on the changes in ionic strength and pH of the solution and not on the concentration of assembled capsids. The long-term stability of capsids in low-concentration conditions in fact enables experiments that could check our predictions.

The concept of line tension energy has been shown to play an important role in the nucleation models of capsid assembly \cite{Zandi, SiberMaj}. While it does not importantly influence the distribution of capsids and protein clusters in the assembled, equilibrium state \cite{SiberMaj}, its presence does lead to the appearance of a barrier in the assembly free energy, resulting in sigmoidal kinetics of capsid assembly, i.e. a characteristic lag time that depends on the barrier height \cite{Zandi}. We have demonstrated that the line tension is necessarily modified by electrostatic effects, and thus the lag time should vary with the salinity and the surface charge density, which is an effect that has only partially been accounted for in the existing capsid nucleation theories \cite{Zandi,Zlot2}. Our study also demonstrates that some care must be exercised when applying the approximate DH results, since these may be unreliable in the regime of surface charge densities and salinities relevant to viruses (see Fig. \ref{fig:linefit}), in accordance with the findings in Ref. \onlinecite{SibRudi1}.

In the case of liquid shells, our model is best applied to rupture of charged and sufficiently thin vesicles -- in that case the surface charge we use corresponds approximately to the monopole surface charge of the vesicle. The model is also useful, however, when the vesicle thickness cannot be neglected on the scale of the screening length, as has already been discussed in the case of viruses (see also Refs. \onlinecite{Ober1} that represent the spherical vesicle as two shells, bilayer of charge). In the simpler variant of the fluid shell model considered by Fromherz, no stable minima corresponding to intermediate shapes occur~\cite{Fromherz1983}. The same is seen from our model, in spite of the fact that in addition we also consider the electrostatic effects -- the only shapes that minimize the free energy are the closed vesicle and the flattened disk. This would lead directly to an all-or-none transition in the  assembly of the shells. Indeed it has been pointed out at various times that e.g. the icosahedral capsid assembly can be construed in analogy to a continuous phase transition \cite{Lorman,Ptytsin}, thus requiring no latent heat in the process of assembly that furthermore proceeds without a nucleation process. 

While Fromherz studied a model of the vesicle that contained the non-electrostatic line tension energy and bending energy contributions, Betterton and Brenner~\cite{Betterton1999} studied the vesicle poration neglecting the bending energy contribution, but accounting (at the DH level) for the electrostatic  and non-electrostatic contributions to the line tension energy. Our approach is more general than the two previous approaches, and contains both of them as limiting cases. 

In the study of Betterton and Brenner, the vesicle was modeled as a flat membrane of constant area and charge and the electrostatics was considered in the DH limit only. For this approximation to hold, the pore radius should be small compared to the curvature of the vesicle. It was shown that in such a model there exists a narrow regime where metastable pores occur~\cite{Betterton1999}. We did not observe these effects neither in the DH approximation nor in the full PB solution, even when $\kappa R \leq 1$, i.e. in the regime where the line tension renormalization concept is of limited use. To stabilize the pore in fluid membrane, mechanisms other than those considered here are needed. These may include adsorption of multivalent ions \cite{Bae} or dynamical stabilization of pores \cite{Moroz}.

Although our primary interest is in stability of charged shells, our study offers interesting insight in their possible modes of growth. If we assume that the spontaneous curvature of the shell is zero (as we did), one may model the shell growth by the addition of subunits to the initially flat patch. Our model predicts that a sufficiently large planar piece of shell material will spontaneously close into a sphere when $p \geq 1$, see Eq. (\ref{eq:our_p}) (the shells of smaller radius may form but these require thermal excitations in order to cross a free energy barrier, see Fig. \ref{fig:fig2}). Again we should emphasize that no intermediate shapes between a flat patch and a complete spherical shell exist. The radius of thus formed spherical shell is
\begin{equation}
R_V \geq \frac{{\cal K}}{\gamma - \gamma_{el}}\;.
\label{eq:shell_rad}
\end{equation}
This inequality shows that the charged shell radius in general depends on its surface charge density and the salinity of the solution. 
We apply Eq. (\ref{eq:shell_rad}) to viruses using $\gamma \sim 2-8$ $k_B T$ \cite{footnote_estimation}, ${\cal K} \sim 60$ $k_B T$ 
\cite{SiberMaj}, $c_0=100$ mM, and $\sigma \sim 0.5$ $e_0/$nm$^2$, which yields $R_V \sim 10 - 30$ nm. This is indeed a range of 
radii typical for viruses. We note here that the electrostatic renormalization of line tension is quite small, only $\sim 0.2$ $k_B T$, 
and has thus a small effect on the radius of the virus. The electrostatic effects become more important in solutions of smaller salinities which in principle enables formation of larger viruses. This mode of growth is different from nucleation models of virus capsids that assume a fixed curvature of the growing shell \cite{Zandi,SiberMaj}, but bears some conceptual similarity to other proposed modes of capsid growth \cite{Lorman}. As in previous nucleation models, it is still difficult to explain the characteristic Caspar-Klug type of order of proteins in a complete capsid that is the hallmark of simple viruses.

\section{Summary and conclusion}\label{secpet}

We have performed a detailed study of the contribution of electrostatic interactions to the overall energetics of an open charged spherical shell immersed in a solution of monovalent salt. The results we obtained can be used to assess the stability of charged vesicles and viruses. In addition, our study also offers some fresh insight regarding the growth of charged shells and predicts a lower limit on the charged shell radius that depends on the physical properties of the shell material as demonstrated in Eq. (\ref{eq:shell_rad}). The alternative description we propose provides a characteristic radius of the charged shell without invoking the spontaneous curvature of the shell material \cite{SiberMaj}.

We have proven by a detailed analysis of the numerical solutions of the PB equation as well as by an analytical solution of the DH approximation that electrostatic effects lead to only two stable shapes of an open spherical shell, without any stable intermediates, that minimize the total free energy. These shapes are the {\sl closed vesicle} and the {\sl flattened disk}. 

Concerning the rupture of vesicles, our study improves on previous ones in several important aspects. When considering electrostatic effects we account for the full spherical geometry and shape morphology of the vesicle. Furthermore, for such a geometry (including a pore) we have been able to analytically solve the problem at the DH level of approximation. This should be of use in a broader context of the electrostatic self-energy of complicated shapes [see e.g Ref.~\onlinecite{French} and references therein] that do not lend themselves to an easy explicit solution of the DH equation but can be analyzed approximately by the methods introduced in this work. Finally, the numerical procedure that we developed enabled us to study the nonlinear effects arising in the full Poisson-Boltzmann approach to the problem by showing that the insight given by the DH solution persists also in the nonlinear PB regime. By combining all of this, we identified possible pitfalls in the previous studies that have almost exclusively treated the similar class of problems at the DH level \cite{Kegel1,Betterton1999,Zandi}. 

\section{Acknowledgments}

A.L.B. acknowledges the financial support by the Slovenian Research Agency under the young researcher grant. A.\v{S}. acknowledges support by the Ministry of Science, Education, and Sports of  Republic of Croatia (Project No. 035-0352828-2837). R.P. acknowledges the financial support by the Slovenian Research Agency under Contract No. P1-0055 (Biophysics of Polymers, Membranes, Gels, Colloids and Cells).

\appendix

\section{Solution of DH equation}\label{sec:soldh}

In order to solve Eq. \ref{eq:helm} in the required geometry, we shall partition the space into the internal and external region of the shell, defined by $r<R$ (region I) and $r>R$ (region II), respectively. Since $K_{l+\frac{1}{2}} (\kappa r) \rightarrow \infty$ when $r \rightarrow 0$ and $I_{l+\frac{1}{2}} (\kappa r) \rightarrow \infty$ when $r \rightarrow \infty$, the solutions inside and outside the shell are given by
\begin{eqnarray}
\varphi_{I} (r, \vartheta) &=& \sum_{l=0}^{\infty} A_l\, i_{l} (\kappa r) P_l (\cos \vartheta)\;, \nonumber \\
\varphi_{II} (r, \vartheta) &=& \sum_{l=0}^{\infty} B_l\, k_{l} (\kappa r) P_l (\cos \vartheta)\;,
\end{eqnarray}
where we have introduced $i_{l}\equiv I_{l+\frac{1}{2}}/\sqrt{r}$ and $k_{l}\equiv K_{l+\frac{1}{2}}/\sqrt{r}$ (see e.g. Ref. \onlinecite{Abram}).

Applying the boundary condition in Eq. (\ref{eq:bndry}) and requiring also that the potential be continuous at $r=R$, we get two equations that relate the expansion coefficients $A_l$ and $B_l$ which enables one to completely solve the problem. 

By using both boundary conditions together with elementary properties of modified spherical Bessel functions~\cite{Abram}, one can derive
\begin{equation}
A_l = \frac{\sigma}{2 \varepsilon\varepsilon_0 \kappa}\,\frac{1}{i_l(\kappa R)}\,
\frac{{\cal F}_3 (l, \vartheta_0)}{{\cal F}_0 (l,\kappa R)} \quad {\rm and} \quad  B_l = \frac{i_{l}(\kappa R)}{k_{l}(\kappa R)} A_l\;,
\label{linsigma}
\end{equation}
where
\begin{equation}
{\cal F}_0(l,\kappa R)\equiv\frac{I_{l+3/2}(\kappa R)}{I_{l+1/2}(\kappa R)}+\frac{K_{l+3/2}(\kappa R)}{K_{l+1/2}(\kappa R)}
\label{firstdef}
\end{equation}
and
\begin{eqnarray}
{\cal F}_3 (l, \vartheta_0) &\equiv& (2l+1) \int_{-1}^{\cos \vartheta_0} P_l (x)\,\mathrm{d}x = \nonumber\\
&=& P_{l+1} (\cos \vartheta_0) - P_{l-1} (\cos \vartheta_0)\;.
\label{seconddef}
\end{eqnarray}
While deriving the above identity we have used well known properties of Legendre polynomials \cite{Abram}. This completes the solution and enables one to write the potential in the whole space, and in particular the potential on the shell surface itself, $\varphi(r = R, \vartheta)$.

\section{(In)validity of the asymptotic expansion of free energy}\label{sec:appz}

From Eq.~(\ref{eq:asymptDH1}) we can deduce the expansion coefficients for the free energy in the limit $\rho\to\infty$\,,
\begin{equation}
\lim_{\rho \rightarrow \infty} F_{el}^{DH} = \frac{\pi R^2\sigma^2}{ \varepsilon\varepsilon_0 \kappa} \frac{1 + \cos\vartheta_0 }{2} + {\cal O} \left( \frac{1}{\rho^2} \right )\;.
\label{eq:asymptDH2}
\end{equation}
Due to the way the modified Bessel functions of the first and third kind appear in the summation, the first order correction in $1/\rho$ cancels out and no $1/\rho$ term is obtained in this limit, but there may be a term in $1/\rho^2$. The former conclusion is consistent with the general analysis of Duplantier \cite{Duplantier} but the latter is not. 

In the limit of a closed sphere, $\vartheta_0\to0$, the asymptotic expansion can be summed exactly, yielding
\begin{equation}
\lim_{\vartheta_0\to0}F\asymp\frac{\pi  R^2 \sigma^2}{ \varepsilon\varepsilon_0 \kappa}\,\left[1+{\cal O}\left(\frac{1}{\rho^3}\right)\right]\;,
\end{equation}
[compare this with Eq. (\ref{eq:DHlimitclosed})]. From this we see that indeed there are no first or second order contributions for the case of a {\sl closed} spherical shell in the limit $\rho\to\infty$, in perfect accordance with Duplantier \cite{Duplantier}.

Since the only term in the free energy with explicit $\rho$ dependence is the one containing the ratios of Bessel functions [${\cal F}_0 (l,\kappa R)$], the validity of the asymptotic expansion is determined by the order of the Bessel functions $l$\,. An estimate of the range of validity of the expansion is given in Ref. \onlinecite{Arfken} in the form
\begin{equation}
8\rho\gg4l^2-1\;.
\end{equation}
Thus, for $l$ much larger than one, the region of validity of the asymptotic expansion can be very far out ($\rho \gg 50$ for $l=10$\,, for instance). 

In addition, the sum over $l$ in the expression for the free energy also includes the $l$-dependent factors contingent on Legendre polynomials of orders $l+1$ and $l-1$ as well as the argument $\cos{\vartheta_0}$, which stems from the angular dependence due to the finite opening of the spherical shell. In principle this part of the free energy still has to be summed to infinity, regardless of where in the $\rho$ regime we are. 

Let us now try putting all this together in a final assessment of the validity of the asymptotic expansion of the free energy. Since we are dealing with a DH type of problem, the sharpness of the edge of the partially opened shell is defined on the scale of the screening length $1/\kappa$. If the sharpness of the edge is on the scale bigger than the screening length, the screening will smoothen out all the effects of the edge. In analytical terms the azimuthal sharpness of the edge must be given by
\begin{equation}
\Delta\vartheta\sim\frac{\kappa^{-1}}{R}=\frac{1}{\rho}
\end{equation}
in order not to be washed away by screening. How many Legendre polynomials ($l$'s) must one sum up in order to get this azimuthal sharpness? About
\begin{equation}
l_{max}\sim\frac{\pi}{\Delta\vartheta}\sim\pi\rho\;.
\end{equation}
Thus as $\rho\to\infty$\,, so does the maximal number of summation terms in the free energy, $l_{max}\to\infty$. This means that when considering the validity of the asymptotic expansion, the $l_{max}$ is {\sl additionally constrained by the fact that it needs to represent the edge correctly}, so that one gets
\begin{equation}
8 \rho\gg 4(\pi\rho)^2-1\;.
\end{equation}
This is the condition for the applicability of the asymptotic expansion to the problem of a partially opened spherical shell that has a hole with a sharp edge. It is clear that this condition can not be satisfied for any value of $\rho$ in the asymptotic regime ($\rho \gg 1$).

\section{DH free energy obtained by summing interactions between screened charges}\label{sec:appa}

Within the range of applicability of the DH theory the free energy of fixed external charges, such as those residing on the incomplete spherical shell, can be calculated via a direct summation of the screened interaction free energies \cite{Verwey}. If the fixed external charge is described with the charge density $\rho_0({\bf r})$ then the interaction energy of the charges is given by
\begin{equation}
F_{el}^{DH} = \frac{1}{2} \int\!\!\!\int \mathrm{d}^3{\bf r}\,\mathrm{d}^3{\bf r}'\;\rho_0({\bf r}) {\cal G}({\bf r} - {\bf r}') \rho_0({\bf r}')\;,
\label{eq:DHdiscrete1}
\end{equation}
where ${\cal G}({\bf r} - {\bf r}')$ is the Green's function of the DH equation in empty space. The equivalence of the forms of Eqs. (\ref{eq:DHdiscrete1}) and (\ref{neweqfree}) is exact as long as there are no dielectric boundaries in the system. If there are, the form of the Green's function is modified but the equivalence of the two forms remains valid.

For point-like charges with $\rho_0({\bf r}) = \sum_i q_i \delta^3({\bf r} - {\bf r}_i)$ this leads to the following form of the electrostatic free energy obtained simply by summing the screened electrostatic interaction of these charges:
\begin{equation}
F_{el}^{DH} = \frac{1}{2} \frac{1}{4 \pi  \varepsilon\varepsilon_0} \sum_{i,j} \frac{q_i q_j 
\exp (-\kappa r_{ij})}{r_{ij}}\;,
\label{eq:DHdiscrete}
\end{equation}
where $r_{ij} = | {\bf r}_i -  {\bf r}_j |$. A marked advantage of the representation in Eq. (\ref{eq:DHdiscrete}) with respect to the form of Eq. (\ref{eq:finalenerg}) is that the effective interaction potential or equivalently the Green's function entering the former is short-ranged, i.e. cut-off on the length scale of $1/\kappa$. 

\begin{figure}[ht]
\centerline{
\epsfig {file=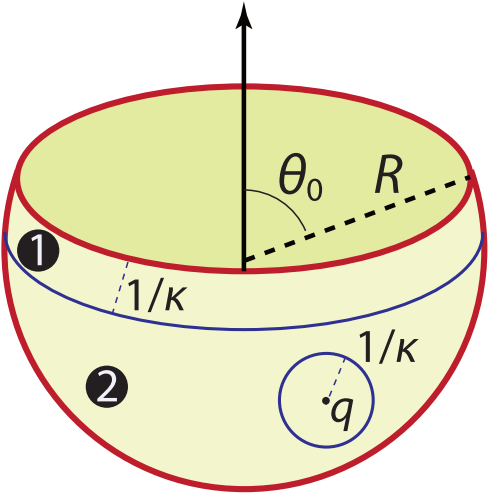,width=6.5cm}
}
\caption{Different regions in an incomplete shells, selected on the criterion of their distance from the edge. An isolated point charge is shown in region 2 with the surrounding that contributes to its electrostatic energy.}
\label{fig:fig_DHderiv}
\end{figure}

To obtain the electrostatic free energy of the incomplete shell in the DH regime, it is convenient to imagine the shell to be populated by the set of point charges of appropriate surface density, so that the surface charge density is $\sigma$\,, and apply the sum in Eq. (\ref{eq:DHdiscrete}). Since the summation is effectively cut on the scale of $\kappa^{-1}$\,, there are two different sets of effective point charges on the capsid -- ones that are close to the shell edge, so that their distance from the edge is smaller than $\kappa^{-1}$ (region denoted by 1 in Fig.~\ref{fig:fig_DHderiv}), and the others that are located farther away from the edge (region denoted by 2 in Fig.~\ref{fig:fig_DHderiv}). 

The charges in region 2 are surrounded by neighbors from all sides, i.e. they are not under the influence of the edge. The charges in region 1, on the other hand, are influenced by the presence of the edge, as it is located within the salt screening distance. Evaluation of Eq.~(\ref{eq:DHdiscrete}) may seem to still pose a problem, since we have to sum up the interactions on the curved surface of the sphere. These summations can be converted to integrals, if we note that in the limit when the screening distance is quite small, $\kappa R\gg1$\,, the important surrounding of a particular point charge is practically flat -- this is due to the fact that the spherical cap is practically a piece of plane when its opening angle is very small. In our case the opening angle of a spherical cap surrounding a particular point charge is $(\kappa R)^{-1}$, and the assumption of planarity is thus satisfied in the limit we are interested in, and that, in particular, applies well to viruses \cite{Kegel1,SibRudi1}.

For the electrostatic energy of a ``patch'' of infinitesimal area $\mathrm{d}A$ in region 2 (``point charge''), one obtains
\begin{equation}
\mathrm{d}F_{el,2}=\frac{\mathrm{d}A}{2}\frac{\sigma^2}{4\pi\varepsilon\varepsilon_0}\,2\pi\!\!\!\int_0^{\infty}\!\!\!\!\!\!\exp(-\kappa r)\mathrm{d}r=\mathrm{d}A\frac{\sigma^2}{4\varepsilon\varepsilon_0\kappa}\;,
\end{equation}
where the assumption of flatness of the surrounding is expected to hold everywhere (since we integrate to $r=\infty$), but this is of no importance 
as $\kappa$ effectively cuts off the integration. The equation above was obtained by summing the screened interaction of a point charge 
(infinitesimal patch) with the rings of infinitesimal width and integrating over all the rings in the plane. Note how the electrostatic energy of the 
patch of area $\mathrm{d}A$ does not depend on its position, which means that the total energy of charges in region 2 is 
\begin{equation}
F_{el,2}=\frac{\sigma^2}{4\varepsilon\varepsilon_0\kappa}A_2\;, 
\end{equation}
where $A_2$ is the area of region 2.

To calculate the electrostatic free energy of region 1 we approximate it as a cylindrical shell with radius of $R\sin\vartheta_0$\,. We introduce the coordinate $z$ measured along the cylindrical shell, going from $z=0$ at the contact of regions $1$ and $2$ to $z=1/\kappa$ at the edge. The relevant infinitesimal area is now $\label{eq:infarea}\mathrm{d}A=2\pi R\sin\vartheta_0\,\mathrm{d}z\;.$ One could introduce a better approximation of region 1 as a conical shell with the infinitesimal area $\mathrm{d}A=2\pi\left(R-z\cos\vartheta_0\right)\sin\vartheta_0\,\mathrm{d}z$\,, but this can obviously be written as a sum of two parts, the first corresponding to the cylindrical shell area above,  and a higher order contribution which is irrelevant for our purposes.

The electrostatic energy of region 1 can now be calculated as follows:
\begin{equation}
F_{el,1} = \frac{2\pi R\sin\vartheta_0\sigma^2}{8 \pi \varepsilon\varepsilon_0} \int_{z=0}^{z=1/\kappa} \mathrm{d}z \int_{S_z} \frac{e^{-\kappa r}}{r} \mathrm{d}S_z\;,
\end{equation}
where $S_z$ now represents the surface obtained by an intersection of a circle, centered on a certain $z$ coordinate and of radius $1 / \kappa$, and 
the cylindrical shell. This surface is a circle without a circular segment of angle $2 \phi_0$, where $\phi_0 = \arccos (1 - \kappa z)$. The integrations over the surface can be written as
\begin{eqnarray}
\int_{S_z} \frac{e^{-\kappa r}}{r} \mathrm{d}S_z &=& 2 \int_{\phi=0}^{\phi=\pi/2} \mathrm{d}\phi \int_{r=0}^{r=\frac{1-z \kappa}{\kappa \cos \phi} } e^{-\kappa r} \mathrm{d}r \nonumber \\
&+& 2 \int_{\phi=\pi/2}^{\phi=\pi} \mathrm{d}\phi \int_{r=0}^{r=\infty } e^{-\kappa r}\mathrm{d}r\;, 
\label{newsetslabel}
\end{eqnarray}
where the limits of integration in the second integral were extended to infinity, in the spirit of the derivation of electrostatic interaction in the region 2. The limits of integration are thus either terminated at the edge [$r_{max}=(1-z \kappa)/(\kappa \cos \phi)$] or extended to infinity wherever possible, i.e. where the local neighborhood is similar to the one in region 2. The second integral ($I_2$) is easy, but the first one ($I_1$) is still nontrivial. We write it down in full, where we take into the account the integration over $z$ as well:
\begin{equation}
I_1 \equiv 2 \int_{0}^{1/\kappa}\mathrm{d}z\int_{0}^{\pi/2}\mathrm{d}\phi \int_{0}^{\frac{1-z \kappa}{\kappa \cos \phi} } e^{-\kappa r}\mathrm{d}r.
\end{equation}
One can interchange the order of integration over $z$ and $\phi$\,, since the integration limits are not interdependent. Carrying out the integration over $r$ by introducing $1-\kappa z=\cos\phi_0$ as a new variable, while defining
\begin{equation}
{\cal B} \equiv\int_{\phi=0}^{\phi=\pi/2}\mathrm{d}\phi\left \{ \cos\phi\left[ 1-\exp \left(- \frac{1}{\cos \phi} \right ) \right] \right \},
\end{equation}
we get
\begin{equation}
I_1=\frac{2}{\kappa^{2}}\left(\frac{\pi}{2}-{\cal B}\right)\;.
\end{equation}
Putting together both integrals in Eq. \ref{newsetslabel} we see that in comparison with the same integral in region 2, which yielded $\pi\kappa^{-2}$\,, the corresponding contribution in region 1 is reduced by a factor of $2{\cal B}/\pi$ because of the effects of the edge.
The electrostatic contribution to the free energy of region 1 is thus
\begin{equation}
F_{el,1}=\frac{2 \pi R \sin \vartheta_0 \sigma^2}{4 \varepsilon \varepsilon_0 \kappa^2} \left( 1 - \frac{{\cal B}}{\pi} \right)\;.
\end{equation}

The total electrostatic energy of the incomplete shell is then 
\begin{equation}
F_{el}=\frac{\sigma^2}{4\varepsilon\varepsilon_0\kappa} A_2 + F_{el,1}\;,
\end{equation}
where $A_2$ is the area of region 2, 
\begin{equation}
A_2 = 2 \pi R^2 (1+ \cos \vartheta_0) - \frac{2\pi R}{\kappa}\sin\vartheta_0\;.
\end{equation}
Therefore, we get for the total electrostatic energy
\begin{equation}
F_{el}=\frac{\pi R^2\sigma^2}{\varepsilon\varepsilon_0 \kappa}\left(\frac{1+\cos\vartheta_0}{2}-\frac{\mathcal{B}}{2 \pi \kappa R}\sin\vartheta_0\right)\quad.
\end{equation}
Numerical integration gives $\mathcal{B}\approx0.726379$\,. We define 
\begin{equation}
f=\frac{\mathcal{B}}{2\pi}\approx0.115607\;, 
\end{equation}
and thus
\begin{equation}
F_{el}=\frac{\pi R^2\sigma^2}{\varepsilon\varepsilon_0 \kappa}\left(\frac{1+\cos\vartheta_0}{2}-\frac{f}{\kappa R}\sin\vartheta_0\right)\;.
\end{equation}


\begin{thebibliography}{99}
\bibitem{footnote_estimation} We have estimated the non-electrostatic part of the line tension similarly as 
in Ref. \onlinecite{Zandi}, on the basis of experiments reported in Ref. \onlinecite{zlotnick_hepB}.
\end{thebibliography}
\end{document}